\documentclass[conference]{IEEEtran}
\IEEEoverridecommandlockouts
\usepackage{amsmath,amssymb,amsfonts}
\usepackage{algorithmic}
\usepackage{graphicx}
\usepackage{textcomp}
\usepackage{xcolor}
\def\BibTeX{{\rm B\kern-.05em{\sc i\kern-.025em b}\kern-.08em
    T\kern-.1667em\lower.7ex\hbox{E}\kern-.125emX}}

\usepackage[ruled,linesnumbered,vlined]{algorithm2e}
\usepackage{subfig}
\usepackage{url}
\usepackage{booktabs}
\usepackage{color, colortbl}
\definecolor{Gray}{gray}{0.9}
\newcommand{\ourimpl}{\textsc{SubGraph2Vec}}
\newcommand{\fasciaimpl}{\textsc{Fascia}}

\usepackage{amsthm}

\begin{document}

\title{SubGraph2Vec: Highly-Vectorized Tree-like Subgraph Counting}

\author{
\IEEEauthorblockN{Langshi Chen}
\IEEEauthorblockA{\textit{Indiana University}\\
lc37@indiana.edu}
\and
\IEEEauthorblockN{Jiayu Li}
\IEEEauthorblockA{\textit{Indiana University}\\
jl145@indiana.edu}
\and
\IEEEauthorblockN{Ariful Azad}
\IEEEauthorblockA{\textit{Indiana University}\\
azad@iu.edu}
\and
\IEEEauthorblockN{Cenk Sahinalp}
\IEEEauthorblockA{\textit{Indiana University}\\
cenksahi@indiana.edu}
\and
\IEEEauthorblockN{Madhav Marathe}
\IEEEauthorblockA{\textit{University of Virginia}\\
marathe@virginia.edu}
\and
\IEEEauthorblockN{Anil Vullikanti}
\IEEEauthorblockA{\textit{University of Virginia}\\
vsakumar@virginia.edu}
\and
\IEEEauthorblockN{Andrey Nikolaev}
\IEEEauthorblockA{\textit{Intel Corporation}\\
andrey.nikolaev@intel.com}
\and
\IEEEauthorblockN{Egor Smirnov}
\IEEEauthorblockA{\textit{Intel Corporation}\\
egor.smirnov@intel.com}
\and
\IEEEauthorblockN{Ruslan Israfilov}
\IEEEauthorblockA{\textit{Intel Corporation}\\
ruslan.israfilov@intel.com}
\and
\IEEEauthorblockN{Judy Qiu}
\IEEEauthorblockA{\textit{Indiana University}\\
xqiu@indiana.edu}
}

\maketitle
\IEEEpubidadjcol

\begin{abstract}
Subgraph counting aims to count occurrences of a template T in a given network G(V, E). It is a powerful graph analysis tool and has found real-world applications in diverse domains. Scaling subgraph counting problems is known to be memory bounded and computationally challenging with exponential complexity. Although scalable parallel algorithms are known for several graph problems such as Triangle Counting and PageRank, this is not common for counting complex subgraphs. Here we address this challenge and study connected acyclic graphs or trees. We propose a novel vectorized subgraph counting algorithm, named \ourimpl{}, as well as both shared memory and distributed implementations: 
1) reducing algorithmic complexity by minimizing neighbor traversal; 
2) achieving a highly-vectorized implementation upon linear algebra kernels to significantly improve performance and hardware utilization.
3) \ourimpl{} improves the overall performance over the state-of-the-art work by orders of magnitude and up to 660x on a single node.
4) \ourimpl{} in distributed mode can scale up the template size to 20 and maintain good strong scalability.
5) enabling portability to both CPU and GPU.

\end{abstract}


\maketitle

\section{Introduction}
\label{sec:introduction}

Counting tree-like subgraphs from a large network is fundamental in graph problems. It has been used in real-world applications across a range of disciplines, including:
\begin{itemize}
    \item Social network analysis: Online social network has billion- or trillion-sized network, where a certain group of users may share specific interests ~\cite{ugander2013subgraph}~\cite{chen:icdm16}.
    
    \item Bioinformatics: The frequency or distribution of the occurrence of each different testing templates may characterize a protein-protein interaction network ~\cite{AlonBiomolecularNetworkMotif2008}~\cite{ SlotaFastApproximateSubgraph2013}, where repeated subgraphs are crucial in understanding cell physiology as well as developing new drugs.  ~\cite{battiston2017multilayer}
    
    \item Computing kernel of other algorithms: Sub-tree counting is one of the computing kernels of bounded treewidth subgraph (such as circles, cactus graphs,  series-parallel graphs etc.) counting problem\cite{chakaravarthy2016subgraph} and also the kernel of network clustering \cite{bordino2008mining}.

\end{itemize}

Despite subgraph counting plays an important role in discovery of patterns in a graph network, counting the exact number of subgraphs of size $k$ in a $n$-vertex network takes $O(n^k)$ time \cite{SlotaFastApproximateSubgraph2013}, which is computationally challenging even for moderate values of $n$ and $k$. In fact, determining whether a graph $G$ contains a subgraph to $H$ is a related graph isomorphic problem that is NP-complete ~\cite{cook1971complexity}. 

Alon et al. \cite{alon_color-coding_1995} provides an approximate algorithm, color-coding, to estimate the number of subgraphs with statistical guarantees. 
Although the color-coding algorithm in~\cite{AlonBiomolecularNetworkMotif2008} has a time complexity linear in network size, it is exponential to subgraph size. 
Therefore, efficient parallel implementations are the only viable way to count subgraphs from large-scale networks. To the best of our knowledge, a multi-threaded implementation named \fasciaimpl{}~\cite{SlotaFastApproximateSubgraph2013} is considered to be the state-of-the-art work in this area. Still, it takes \fasciaimpl{} more than 4 days (105 hours) to count a 17-vertex subgraph from the RMAT-1M network (1M vertices, 200M edges) on a 48-core Intel (R) Skylake processor. While our proposed algorithm named \textbf{\ourimpl{}} takes only 9.5 minutes to complete the same task on the same hardware. \par

The primary contributions of this paper are as follows:
\begin{itemize}
\item {\bf Algorithmic Design.} We identify and reduce the computation complexity of the sequential color-coding algorithm, which also helps reduce communication overhead in distributed systems.
\item {\bf System design and optimization.} We design a data structure as well as a thread execution model to leverage the hardware efficiency of using linear-algebra kernels in terms of vector processing units (VPU) and memory bandwidth.
\item {\bf Portability to the distributed system and GPU} We scale out our single node implementation on a distributed system with near-linear strong scalability. In addition, we export the codes to NVIDIA GPU by using NVIDIA cuSPARSE kernels thanks to our modular system design.  
\end{itemize}
The codebase of our work on \ourimpl{} is made public in our open-sourced repository \cite{sub2vec}.
\section{Preliminaries}
\label{sec:preliminary}

\subsection{Motivation}
\begin{figure}[ht]
    \centering
\includegraphics[width=0.8\linewidth]{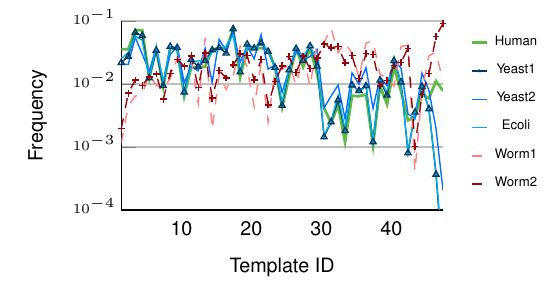}    
\caption{A comparison of treelet distributions of five PPIN networks by \ourimpl{}}
    \label{fig:ppin}
\end{figure}

Counting repeated subgraphs (motifs) can be used to measure topological features and further reveal the similarity of any given two networks. Fig.~\ref{fig:ppin} illustrates such a real-world application in protein-protein interaction network (PPIN), where we use \ourimpl{} to count tree-like motifs in Fig.~\ref{fig:templates}(a) to estimate their frequencies. We compared the PPI networks of humans\cite{hodzic2019combinatorial}, yeasts, C. elegans\cite{Yeast}, and E. coli\cite{EColiNet}, and we have findings on 
the normalized treelet distributions for the unicellular organisms: Ecoli and yeasts are very close, while the more complex C. elegans (a kind of worm) is significantly different. \par
As PPINs usually include many false (positive and negative) and missing interactions\cite{AlonBiomolecularNetworkMotif2008} in practice, an occurrence of a specific treelet may include additional (or missing) edges in different networks. Counting non-induced subgraph is more suitable to obtain reliable and robust results\cite{AlonBiomolecularNetworkMotif2008}.  
In Fig.~\ref{fig:templates}(a), we show 47 unlabeled treelets with similarity. They have the same size of 9 vertices but vary slightly in topology.  Note that an induced subgraph of a graph G(V,E) is a subset of the vertices of the graph G(V,E) as well as with any edges connecting pairs of vertices in that subset. There are many more non-induced subgraphs isomorphic to a given topology as they allow missing edges among vertex pairs. Thus, it is challenging to count non-induced subgraphs of a network.  

\subsection{Statement of Problem}
\label{sub:statement_problem}
\subsubsection{Subgraph Counting}
Subgraph finding and counting is a widely studied subject. A (non-induced) subgraph of a simple unweighted graph $G(V,E)$ is a graph $H(V_H,E_H)$ satisfying $V_H \subset V$ and $E_H \subset E$. $H$ is an embedding of a template graph $T$ if $T$ is isomorphic to $H$. The subgraph counting problem is to count the number of all embeddings of a given template $T$ in a network $G$. We use $emb(T,G)$ to denote the number of all embeddings of template $T$ in network $G$.

\subsubsection{Color coding}
Color coding \cite{alon_color-coding_1995} is an algorithmic technique to discover network motifs. 
Given a $k$-node template $T$, it assigns random colors between $0$ and $k-1$ to each vertex of a network graph $G$, and it counts the number of the occurrences of colorful embedding, which is isomorphic to $T$ while having distinct colors on each vertex. 
Both theoretical proof \cite{alon_color-coding_1995, alon2008biomolecular, chakaravarthy2016subgraph} and experiments \cite{AlonBiomolecularNetworkMotif2008, slota_parallel_2015} show that,  with proper normalization, the count of colorful embeddings is an unbiased estimator of the actual count of embeddings. Alon et al. \cite{alon_color-coding_1995} proved a guarantee of bounding the count by $(1\pm \epsilon)emb(T,G) $ with a probability of $1-2\delta$ after running at most $N$ iterations of the algorithm.
 \begin{figure}[ht]
     \centering
     \includegraphics[width=0.85\linewidth]{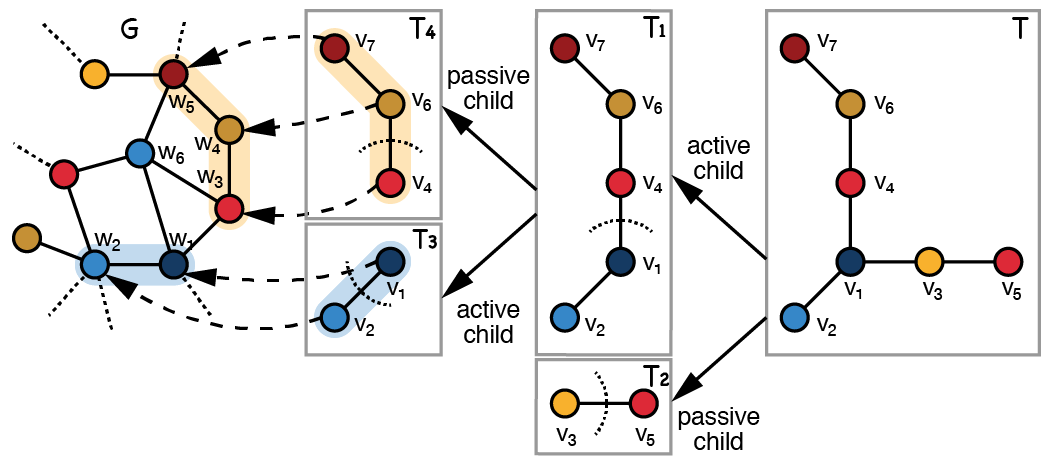}
     \caption{Illustration of the template partitioning within a
     colored input $G=(V,E)$}
     \label{fig:partitionTemplates}
 \end{figure}
\subsection{Counting Tree-like Subgraph}
By applying color coding, \cite{AlonBiomolecularNetworkMotif2008} provides a \emph{fixed parameter tractable} algorithm to address the subgraph counting problem where $T$ is a tree.
It has a time complexity of $O(c^k\text{poly}(n))$, which
is exponential to template size $k$ but polynomial to vertex number $n$. Algorithm~\ref{alg:sequential}
describes the standard sequential algorithm with definition and notation shown in Table~\ref{tab:notation}, which contains three important steps as follows.
\begin{algorithm}[ht]
\small
\caption{Standard Sequential Algorithm}
\label{alg:sequential}
\SetKwInOut{Input}{input}
\SetKwInOut{Output}{output}
$N=O(\frac{e^k\log(1/\delta)}{\epsilon^2})$ \tcp*[h]{required iterations to converge}\\
Partition $T$ into sub-templates $T_s$ \\ 
\For{$j=1$ to $N$}
{
    color all $V_i \in G(V,E)$  randomly \\
    \colorbox{gray!30}{counting $T_s$ in a dynamic programming procedure} \\
    $P \gets$ probability that the template is colorful\\
    $\alpha \gets$ number of automorphisms of $T_0$\\
    $finalCount[j] \gets \frac{1}{P\alpha}\sum_{i} \sum_{C} \mathbf{M}_{0}(i,I_C)$ 
}
Output the average of all $finalCount$.
\end{algorithm}
\subsubsection{Random Coloring}
Each vertex $v \in G(V,E)$
is given an integer value of color randomly selected
between $0$ and $k-1$, where $k \ge |V_T|$ (we consider $k=|V_T|$ for simplicity). $G(V,E)$ is therefore converted to a labeled graph.
\subsubsection{Template Partitioning}
For tree-like templates, we can recursively partition $T$ into a chain of sub-templates $T_s$ until
the sub-template containing only one vertex 
When partitioning a template $T$, a single vertex $\rho$ is selected as the root 
while $T_s(\rho)$ refers to the $s$-th sub-template rooted at $\rho$. Secondly, 
one of the edges $(\rho, \tau)$ adjacent to root $\rho$ is cut, creating two child sub-templates. 
The child holding $\rho$ as its root is named \emph{active child} and denoted as $T_{s,a}$. 
The child rooted at $\tau$ of the cutting edge is named \emph{passive child} and 
denoted as $T_{s,p}$.  

\subsubsection{Dynamic Programming} 
\begin{algorithm}[ht]
\small
\caption{Dynamic Programming in Standard Sequential Algorithm}
\label{alg:seq-dp}
\SetKwInOut{Input}{input}
\SetKwInOut{Output}{output}
\Input{$G(V,E), T$}
\Output{$\mathbf{M}_{s}$}
\ForAll{sub-templates $T_s$ in reverse order of their partitioning}
{
    \eIf{$T_s$ consists of a single vertex}
    {
        \ForAll{$V_i \in V$}
        {
            $\mathbf{M}_s(i,\text{color of  $V_i$}) \gets 1$
        }
    }
    {\tcp*[h]{$T_s$ has an active child $T_{s,a}$ and a passive child $T_{s,p}$}\\
        \ForAll{vertices $V_i \in V$} 
        {
            \ForAll{color sets $C_s$ satisfying  $|C|=|T_s|$ }
            {
              \ForAll{color sets $C_{s,a}$ and $C_{s,p}$ created by splitting  $C_s$ satisfying  $|C_{s,a}|=|T_{s,a}|$ and $|C_{s,p}|=|T_{s,p}|$}
              {{$\mathbf{M}_s(i,I_s) \gets\sum\limits_{V_j \in N(V_i)} \mathbf{M}_{s,a}(i,I_{s,a})\mathbf{M}_{s,p}(j,I_{s,p})$}}
            }
        }
    }
}
\end{algorithm}
Algorithm~\ref{alg:seq-dp} 
describes the dynamic programming procedure to count partitioned 
template $T$ from the randomly colored $G(V,E)$. 
For bottom sub-template $|T_s|=1$, $M_{s}(i, I_s)$ is $1$ only if $C_s$ equals the color randomly assigned to $V_i$, and otherwise it is $0$. For non-bottom cases where $|T_s|>1$, we obtain 
$M_{s}(i,I_s)$ by multiplying the count values from its two children, which have been calculated in previous steps of dynamic programming. \par

\begin{table}[ht]
\renewcommand\arraystretch{0.8}
    \centering
    \caption{Definitions and Notations}
    \label{tab:notation}
\setlength{\tabcolsep}{0.5mm}{
    \begin{tabular}{ll}
    \toprule
     Notation & Definition  \\
    \midrule
    $G(V,E)$ or $G$  & The input network \\ 
    $\mathbf{A}_G$ & $|V|\times |V|$ sparse adjacency matrix of $G(V,E)$ \\
    $N(V_i)$ or $N(i)$ & Neighbors of vertex $V_i$\\
    $T$, $T_s$ & The input template and the $s$-th sub-template \\
    $|T_s|$ & Number of vertices in $T_s$ \\
    $T_{s,a}$, $T_{s,p}$ & Active and passive child of $T_s$ \\
    $n$ & $n=|V|$ is the number of vertices in $G$ \\
    $k$ &  $k=|V_T|$ is the number of vertices in $T$ \\
    $C_s$ & Color set for $T_s$ \\ 
    $\mathbf{M}_s$  & $|V|\times\binom{k}{|T_s|}$ dense matrix to store counts for $T_s$ \\
    $\mathbf{M}_{s,a}$, $\mathbf{M}_{s,p}$  & Dense matrix to store counts for $T_{s,a}$, $T_{s,p}$ \\
    $\mathbf{B}$  & $\mathbf{B}=\mathbf{A}_G \mathbf{M}_{s,p}$, the sum of the counts of all neighbors. \\
    $I_{C_s}$ or, $I_{s}$ & Column index of color set $C_s$ \\ 
    \bottomrule
    \end{tabular}
}
\end{table}
\section{Related Work}
\label{sub:related_work}
The graphlet frequency distance was proposed by Przulj et. al\cite{prvzulj2004modeling} as a global comparative measure based on the local structural characteristics of different networks. Bordino et al. \cite{bordino2008mining} demonstrates that one can use the relative frequency of subgraphs within networks to distinguish and cluster different networks. \par 

A tree subgraph enumeration algorithm by combining color coding with a stream-based cover decomposition was developed in ~\cite{zhao2010subgraph}. To process massive networks, \cite{zhao_sahad:_2012} developed a distributed color-coding based tree counting solution upon the MapReduce framework in Hadoop, \cite{SlotaComplexNetworkAnalysis2014a} implemented an MPI-based solution, and ~\cite{zhao2018finding}~\cite{chen2018high} pushed the limit of subgraph counting to process billion-edged networks and trees up to 15 vertices. \cite{guelsoy2012topac} developed a coloring method that achieves a provable confidence value in a small number of iterations. \par

\cite{alon_color-coding_1995} proved that color-coding could apply to subgraph counting problems, where the template is a tree, a cycle, or any graph with bounded treewidth. 
\cite{chakaravarthy2016subgraph} is a color-coding implementation applying to all templates with a treewidth of no more than 2. Beyond counting trees, a sampling and random-walk based technique \cite{AhmedEfficientGraphletCounting2015}\cite{chen:icdm16} could count graphlets, small induced graph with size up to 4 or 5.\par

Other subgraph topics include: 1) \emph{subgraph finding}. As in~\cite{EkanayakeMIDASMultilinearDetection2018}, paths and trees with size up to 18 could be detected by using multilinear detection; 2) \emph{Graphlet Frequency Distribution} estimates relative frequency among all subgraphs with the same size~\cite{PrzuljBiologicalNetworkComparison2010}; 3)
\emph{clustering} networks by using the relative frequency of their subgraphs
~\cite{RahmanGraftEfficientGraphlet2014}. \emph{Subgraph Matching} finds
and enumerates all isomorphic subgraphs to a given template from input network. \cite{sun2012efficient} contributes 
an online algorithm to query subgraph templates from 
billion-node network by using 
intelligent graph exploration to 
replace expensive join operations. \cite{lee2012depth}
compares and summarizes subgraph isomorphism algorithms 
in graph databases. Later on ~\cite{bi2016efficient} 
improves the performance of subgraph matching up to three
orders of magnitude by postponing the 
Cartesian products based on the structure of a query to minimize the redundant Cartesian products. 
\cite{RezaPruneJuicePruningTrillionedge2018, reza2017towards} provides a pruning method on labeled networks and graphlets to reduce the vertex number by orders of magnitude prior to the actual counting.\par

\section{Algorithmic Design of \ourimpl{}}
\label{sec:alg-design}

Unlike standard sequential algorithm in Algorithm~\ref{alg:seq-dp}, we decouple the dynamic programming into two stages, as shown in Algorithm~\ref{alg:prune-redundancy}: 1) A vertex-neighbour traversal stage, and 2) A counts updating stage. This design brings a two-fold benefit. First, it helps us identify and remove redundant computation and thus reduce the complexity of Algorithm~\ref{alg:seq-dp}, and secondly it enables us to apply the optimization on system design. \par 

\begin{figure}[ht]
    \centering
    \includegraphics[width=0.9\linewidth]{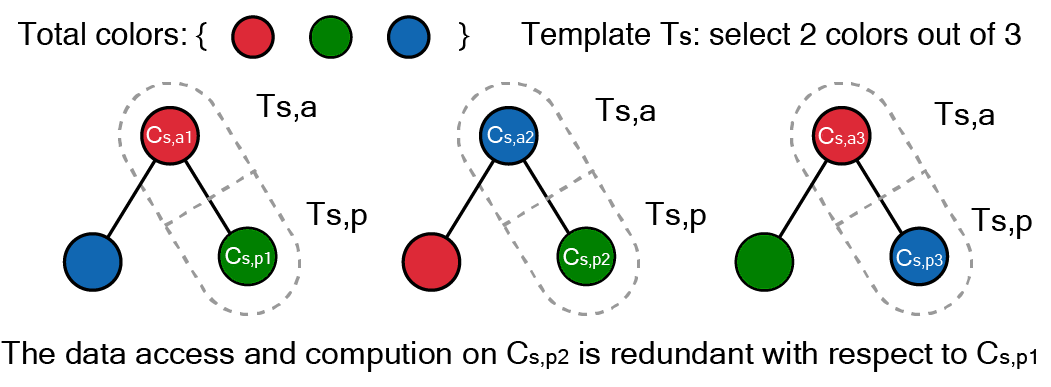}
    \caption{Identify the redundancy of standard color coding in a two-vertex sub-template $T_s$.}
    \label{fig:baseline-redundancy}
\end{figure}

In Algorithm~\ref{alg:seq-dp}, 
it requires $\binom{|T|}{|T_s|}\binom{|T_s|}{|T_{s,a}|}$
times of vertex neighbor traversal for each $V_i$ from line 7 to 9.
However, we find that redundancy of traversal exists as shown in 
Figure~\ref{fig:baseline-redundancy}, where it 
\begin{figure}[ht]
    \centering
    \includegraphics[width=0.85\linewidth]{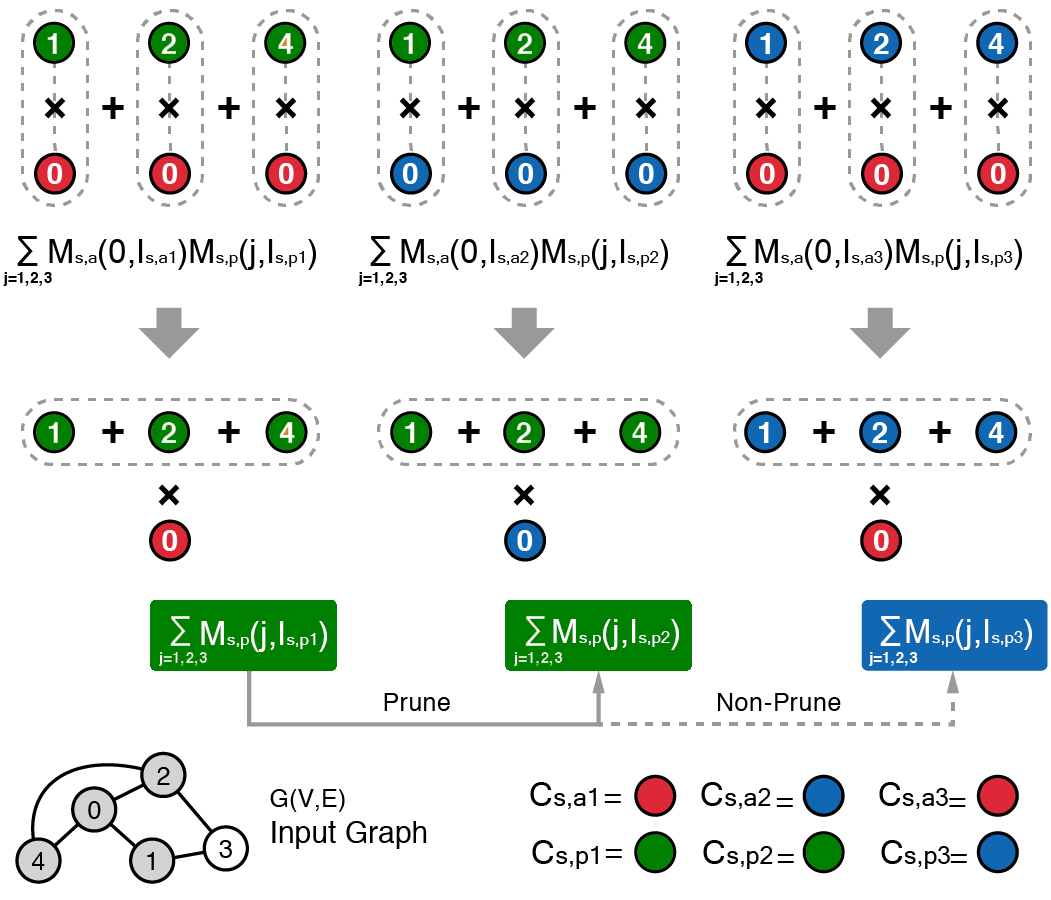}
    \caption{Decouple the vertex neighbor traversal from updating of the count value according
    to distributive property of addition and multiplication in Equation~\ref{eq:transform_traversal}.
    }
    \label{fig:decouple-cache-traversal}
\end{figure}
counts a two-vertex sub-template with a total of three colors. The left case and the middle case have 
the same color set (the same $I_{s,p}$) assigned to their passive child $T_{s,p}$, which 
causes redundant access to $\mathbf{M}_{s,p}(j, I_{s,p})$ when traversing neighbor 
vertices. \par
\begin{algorithm}[ht]
\small
\caption{Dynamic Programming in \ourimpl{}}
\label{alg:prune-redundancy}
\SetKwInOut{Input}{input}
\SetKwInOut{Output}{output}
\Input{$G(V,E), \mathbf{M}_{s,p}, T_s$}
\Output{$\mathbf{M}_{s}$: matrix storing traversal results}
\For{$V_i \in G(V,E)$}
{
        \For{color sets $C_{s,p}$ satisfying $|C_{s,p}|=|T_{s,p}|$}
        {
            \ForAll{$V_j \in N(V_i)$}
            {
               \colorbox{gray!30}{ $\mathbf{B}(i, I_{s,p}) \gets \mathbf{B}(i, I_{s,p}) + \mathbf{M}_{s,p}(j, I_{s,p})$} \\
            }
        }
}
\For{$V_i \in G(V,E)$}
{
    \For{color set $C_s$ satisfying $|C_s|=|T_s|$} 
    {
        \For{color sets $C_{s,a}$, $C_{s,p}$ split from $C_s$}
        {
            \colorbox{gray!30}{$\mathbf{M}_s(i, I_s) \gets \mathbf{M}_s(i, I_s) + \mathbf{M}_{s,a}(i, I_{s,a})\mathbf{B}(i, I_{s,p})$}
        }
    }
}
\end{algorithm}
On the contrary, \ourimpl{} proposes a novel way to accomplish the vertex neighbor traversal 
described from 1 to 4 of Algorithm~\ref{alg:prune-redundancy}:
\begin{enumerate}
    \item The vertex neighbor traversal is decoupled from line 9 of Algorithm~\ref{alg:seq-dp}. 
    \item Only $\binom{|T|}{|T_{s,p}}$ times of traversal is applied on each vertex.
\end{enumerate}
According to distributive property of addition and multiplication, line 9 of Algorithm~\ref{alg:seq-dp} can be re-written as 
\begin{equation}
\label{eq:transform_traversal}
\begin{aligned}
    \sum_{V_j \in N(i)}\mathbf{M}_{s,a}(i, I_{s,a})\mathbf{M}_{s,p}(j, I_{s,p}) \\
    = \mathbf{M}_{s,a}(i, I_{s,a})\sum_{V_j \in N(i)}
    \mathbf{M}_{s,p}(j, I_{s,p})
\end{aligned}
\end{equation}
, where the first item $\mathbf{M}_{s,a}(i, I_{s,a})$ at right-hand only contains count values 
of $V_i$ while the second item  $\sum_{V_j \in N(i)}\mathbf{M}_{s,p}(j, I_{s,p})$ only 
involves traversing neighbors of $V_i$. This decoupled design enables a caching and re-using
of the traversal results (the summation of $\mathbf{M}_{s,p}(j, I_{s,p})$ shown in Figure~\ref{fig:decouple-cache-traversal}), which allows us to reduce the traversal times.\par
The second stage of \ourimpl{} is to update the count values by 
multiplying $\mathbf{M}_{s,a}(i, I_{s,a})$ and $\mathbf{B}(i, I_{s,p})$, and both of them are local to vertex $i$, which improves data locality and allows a vectorized computation when compared to the standard sequential Algorithm~\ref{alg:seq-dp} where the non-consecutive indices of $i$ and $j$ at line 9 do not meet the requirement of SIMD paradigm.  

\section{System Design of \ourimpl{}}
\label{sec:sys-design}

We propose a new scheme in \ourimpl{} to achieve two goals: 1) Vectorize \cite{Vectorization} both stages in Section~\ref{sec:alg-design}. 2) Transform and leverage the vectorized codes into BLAS kernels. 
\subsection{Vectorize Vertex-Neighbor Traversal}
\label{sub:vec-vertex-traversal}

\begin{figure*}[ht]
    \centering
    \includegraphics[width=0.8\linewidth]{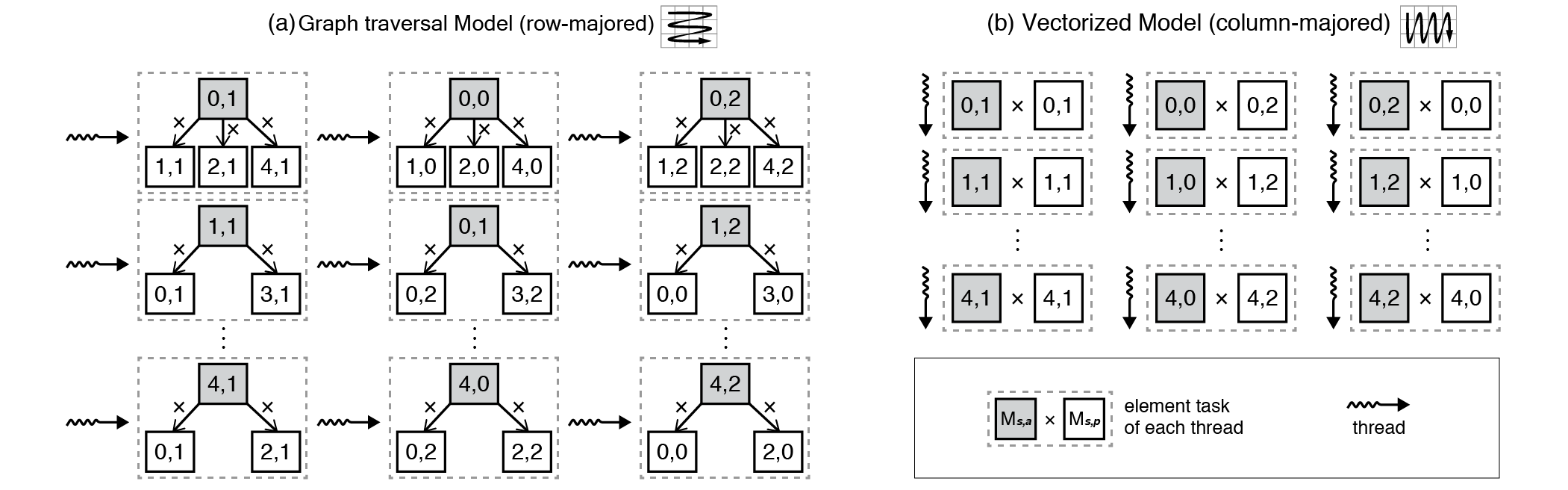}
    \caption{Comparing the thread execution order, where
    (a) Graph traversal Model counts data stored in memory with a row-majored layout, and (b) \ourimpl{} (Vectorized Model) counts data stored in memory with a column-majored
    layout.}
    \label{fig:thd-execution-order}
\end{figure*}
\ourimpl{} utilizes an adjacency matrix, notated as $\mathbf{A}_G$, to store input network $G(V,E)$ in Sparse Row Compressed (CSR) format. $\mathbf{A}_G$ is a sparse 0-1 matrix  satisfying $\mathbf{A}_G(i,j) = 1$ if and only if $V_j \in N(V_i)$.  
Correspondingly, we re-write line 1 to 4 of Algorithm~\ref{alg:prune-redundancy} by  
Algorithm~\ref{alg:prune-adj-mat}, 
where for each $I_{s,p}$, we schedule loops of $V_i$ to threads 
while each thread is vectorizing its own work.
\begin{algorithm}[ht]
\small
\caption{Vectorized Vertex-Neighbor Traversal in \ourimpl{}}
\label{alg:prune-adj-mat}
\SetKwInOut{Input}{input}
\SetKwInOut{Output}{output}
\Input{$\mathbf{A_G}, T_s, \mathbf{M}_{s,p}$}
\Output{$\mathbf{B}$}
\ForAll{color sets $C_{s,p}$ satisfying $|C_{s,p}|=|T_{s,p}|$}
{
    \ForAll(\tcp*[h]{loop is scheduled to threads}){$V_i \in \mathbf{A_G}$}
    {
        \ForAll{$j=\mathbf{A_G}.rowIdx[i]$ to $\mathbf{A_G}.rowIndex[i+1]$}
        {
            \tcp*[h]{thread workload is vectorized} \\
            {$\mathbf{B}(i,I_{s,p}) \gets \mathbf{B}(i,I_{s,p})+ 
            \mathbf{A_G}.val[j]\mathbf{M}_{s,p}(\mathbf{A_G}.colIdx[j], I_{s,p})$}
        }
    }
}
\end{algorithm}
We observe that $j$ has successive values from $A_G.rowIdx[i]$ to 
$\mathbf{A}_G.rowIdx[i+1]$ resulting in coalesced data access to three
dense arrays of $A_G$.
Unfortunately, $\mathbf{A}_G.colIdx[j]$ does not guarantee successive values due to the sparsity of $\mathbf{A}_G$. 
However, advanced compilers
still provide partial vectorization support to this indexed
access pattern. We will introduce our customization in addressing this partial vectorization issue in Section~\ref{sec:use-kernels}.

\subsection{Vectorize Count Updating}
\label{sub:vec-count-update}

In Algorithm~\ref{alg:prune-redundancy}, counting templates at line 8 
cannot be vectorized because the indices are not successive. 
To address this issue, we propose a new scheme illustrated in 
Figure~\ref{fig:thd-execution-order}.
\begin{enumerate}
    \item Change memory layout from row-majored order to column-majored order.
    \item All threads are working on same columns of $\mathbf{M}_s, \mathbf{M}_{s,a}, \mathbf{M}_{s,p}$ 
    concurrently and processing the matrices column by column.
    \item All rows of a column are evenly distributed to threads.
    \item Each thread vectorizes the work on its own portion of rows.
\end{enumerate}
Compared to Algorithm~\ref{alg:seq-dp}, the length of vectorization is converted from $\binom{|T|}{|T_s|}$ to
the million-level number of vertices in $G(V,E)$, which is sufficient to utilize the hardware fully and invariant to different sub-templates. Furthermore, the stride one regular memory access is efficient in prefetching data from memory to cache lines. 

\subsection{Invocation of Linear Algebra Kernels}
\label{sec:use-kernels}
Compared to the graph traversal model, \ourimpl{} is designed to be portable among hardware platforms while keeping high-performance. The vectorized vertex neighbor traversal module in Algorithm~\ref{alg:prune-adj-mat} mathematically equals to an operation of sparse matrix dense vector multiplication (SpMV), which is an essential sparse linear solver on different hardware platforms. 
Correspondingly, line 8 of Algorithm~\ref{alg:prune-redundancy} equals an 
element-wised multiplication and addition of dense vectors (eMA). 
A complete \ourimpl{} made of SpMV and eMA kernels is described in Algorithm~\ref{alg:ourimpl-spmv-ema}, which also applies in-place storage of the SpMV results from the column vector buffer $\mathbf{B}$ back to $\mathbf{M}_{s,p}$ to reduce the memory footprint. \par

To achieve better kernel performance than by using public libraries, we customize both of 
SpMV and eMA kernels. For SpMV, we combine a bundle of SpMV operations in Algorithm~\ref{alg:ourimpl-spmv-ema} into 
a Sparse matrix dense matrix (SpMM) operation shown in Algorithm~\ref{alg:csc-split-spmm}, 
where the right-hand dense
matrix is $\mathbf{M}_{s,p}$. To save peak memory utilization, we also split columns of $\mathbf{M}_{s,p}$ into batches with pre-selected batch size. To improve the load balancing and data locality, we utilize a split compressed sparse column (CSC-Split) format instead of the default compressed sparse row (CSR) format that is widely used by public libraries. CSC-Split format converts the standard CSC format into a fixed number of partitions. Entries of CSC matrix are distributed
to a partition when their row IDs fit into a pre-defined range of that partition. 
Inside a partition, the entries are ordered by their column IDs of CSC format, and therefore
entries sharing the same column ID and adjacent row IDs are bundled together to improve the 
data locality and cache usage. Meanwhile, we store batches of right-hand vectors from $\mathbf{M}_{s,p}$ in a row-majored memory layout, and we set up the batch 
size to the maximal concurrent element number of the hardware SIMD unit. Finally, when a partition
is assigned to a thread, the thread processes its entries one by one while vectorizing the 
computation work on a batch of row entries from $\mathbf{M}_{s,p}$.
\begin{algorithm}[ht]
\small
\caption{\ourimpl{} with Linear Algebra Kernels}
\label{alg:ourimpl-spmv-ema}
\SetKwInOut{Input}{input}
\SetKwInOut{Output}{output}
\Input{$\mathbf{A}_G, T, \epsilon, \delta$}
\Output{A $(\epsilon,\delta)$-approximation to $emb(T, G)$}
$N=O(\frac{e^k\log(1/\delta)}{\epsilon^2})$ \tcp*[h]{required iterations to converge}\\
Partition $T$ into sub-templates $T_s$\\
\For{$j=1$ to $N$}
{\ForAll{$V_i \in G(V,E)$}
    {
        Color $V_i$ by a value randomly drawn from $1$ to $k=|T|$ 
    }
    \For{$s=0, 1, \dots, S-1$}
    {\ForAll{color sets $C_{s,p}$ satisfying $|C_{s,p}|=|T_{s,p}|$}
        {
            \colorbox{gray!30}{$\mathbf{B}\gets \mathbf{A}_G \mathbf{M}_{s,p}(:,I_{s,p})$} \tcp{SpMV kernel}
            $\mathbf{M}_{s,p}(:,I_{s,p}) \gets \mathbf{B}$ \tcp{Sum of neighbor counts}
        }
        \ForAll{color sets $C_s$ satisfying  $|C_s|=|T_s|$}
        {
            $\mathbf{M}_{s}(:,I_{s}) \gets 0$ \\
            \ForAll{color sets $C_{s,a}$  and $C_{s,p}$, created by splitting $C_s$  satisfying  $|C_{s,a}|=|T_{s,a}|$ and $|C_{s,p}|=|T_{s,p}|$}
            {
                {$\mathbf{M}_{s}(:,I_{s}) \gets \mathbf{M}_{s}(:,I_{s}) + \mathbf{M}_{s,a}(:, I_{s,a})\odot \mathbf{M}_{s,p}(:,I_{s,p})$}
                \tcp{eMA kernel}
            }
        }
    }
    $P\gets$ probability that the template is colorful\\
$\alpha \gets$ number of automorphisms of $T_0$\\
$finalCount[j] \gets \frac{1}{P\alpha}\sum_{i} \sum_{C} \mathbf{M}_{0}(i,I_C)$ 
}
Output the average of all $finalCount$.
\end{algorithm}

\begin{algorithm}[ht]
\small
\caption{CSCSplit SpMM for sub-template $T_s$ in \ourimpl{}}
\label{alg:csc-split-spmm}
\SetKwInOut{Input}{input}
\SetKwInOut{Output}{output}
\Input{$A_G, \mathbf{M}_{s,p}, sIdx, BSize, SplitPars$}
\Output{Out}
\ForAll(\tcp*[h]{partition per thread}){$Par \in SplitPars$ }
{
    \ForAll(\tcp*[h]{workload per thread}){$e \in Par$ }
    {
       \For{$j=sIdx, \dots, sIdx+BSize$} 
       {
            $Out(e.rowId,j) \gets Out(e.rowId, j) + \mathbf{A}_G(e.rowId, e.colId)\mathbf{M}_{s,p}(e.colId, j)$ \tcp*[h]{rowId, colId are row and column 
            indices in CSC-Split compressed sparse format}
       }
    }
}
\end{algorithm}
To customize the eMA kernel, we utilize Intel (R) AVX intrinsics, where multiplication and addition are implemented by using the fused multiply-add (FMA) instruction, which cuts the computation instructions by half. 
In addition, there are already substantial research work in developing high-performance linear algebra kernels. The invocation of linear algebra kernels in \ourimpl{}
benefits from: 
1) using formats and kernel implementations tailored for different input datasets;
2) increasingly improved kernel performance on various hardware platforms. 

\section{Experiments and Results}
\label{sec:experiments}
\subsection{Datasets and Templates}
\label{sub:datasets_and_templates}
\begin{figure}[hb]
    \centering
    \subfloat[For protein-protein interaction networks comparison  ]{\includegraphics[width=0.9\linewidth]{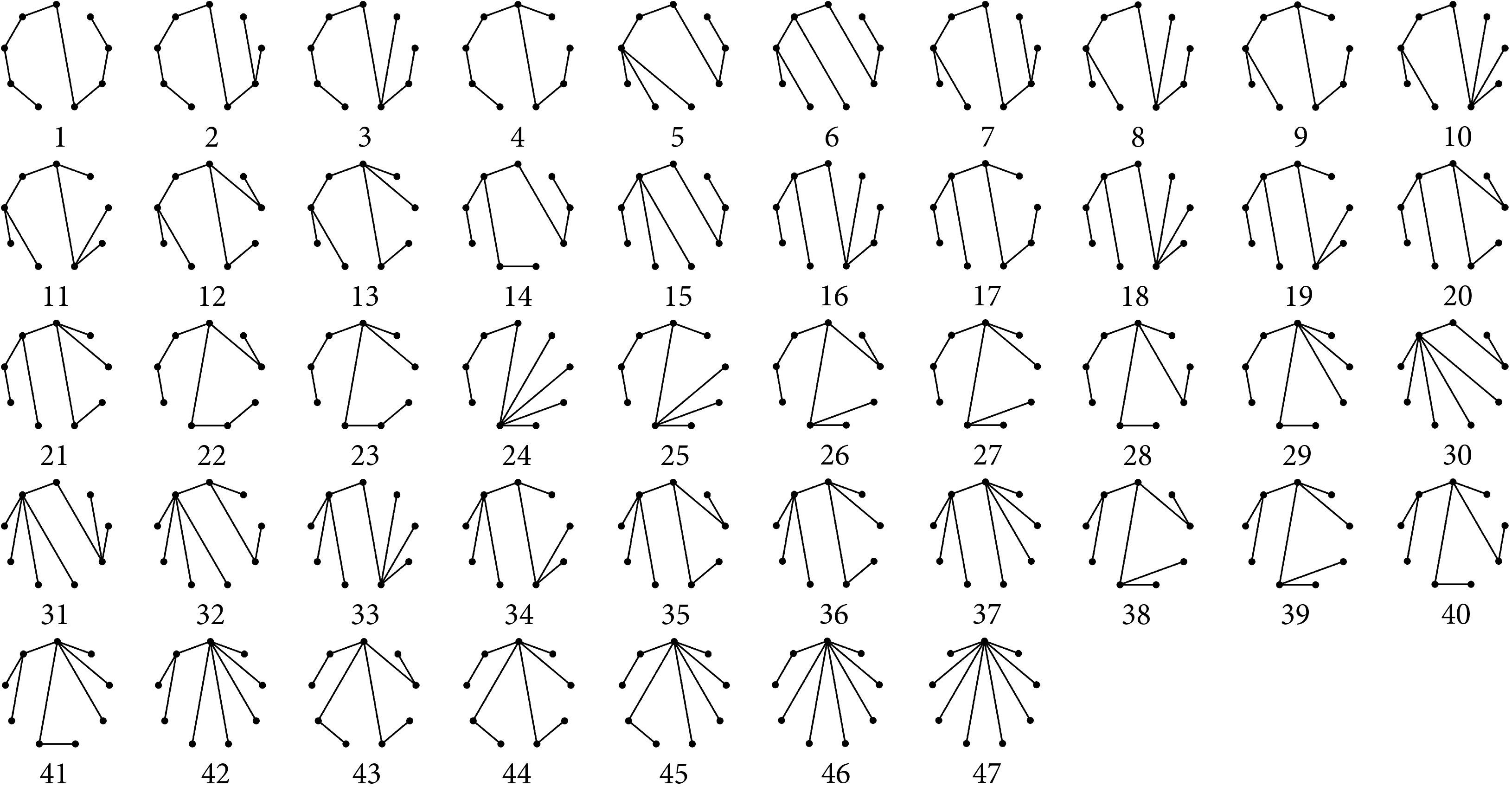}}
    \vfill
    \subfloat[For Social networks, Graph500 and Synthetic data comparison]{\includegraphics[width=0.9\linewidth]{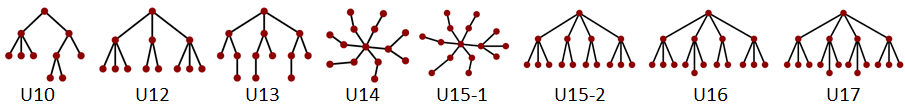}}
    \caption{Templates used in the experiments}
    \label{fig:templates}
    \vspace{-2ex}
\end{figure}

The datasets in our experiments are listed in Table~\ref{tab:datasets}, where \emph{E.coli}~\cite{EColiNet}, \emph{Worm} ~\cite{Yeast} and \emph{Yeast}~\cite{Yeast} are from Biology; \emph{Graph500 Scale=20, 21, 22} are collected from~\cite{graphchallengemit}; \emph{Miami}, \emph{Orkut}, and~\emph{NYC} 
are from ~\cite{barrett_generation_2009}~\cite{snapnets}~\cite{yang_defining_2012}; RMAT are widely used synthetic datasets generated by the RMAT model~\cite{chakrabarti_r-mat:_2004}, 
where we increase parameter $K$ to generate datasets with increasingly skewed degree distribution. Figure \ref{fig:templates}(a) shows all the templates with 9 nodes. The template in Figure \ref{fig:templates}(b) is from \cite{SlotaFastApproximateSubgraph2013}, and the templates with more than 12 nodes are randomly selected. The script used to generate the templates can be found in our open-sourced repository \cite{sub2vec}.

\subsection{Hardware and Software}
\label{sub:exp:impl}

\begin{figure*}[ht]
    \centering
    \includegraphics[width=0.9\linewidth]{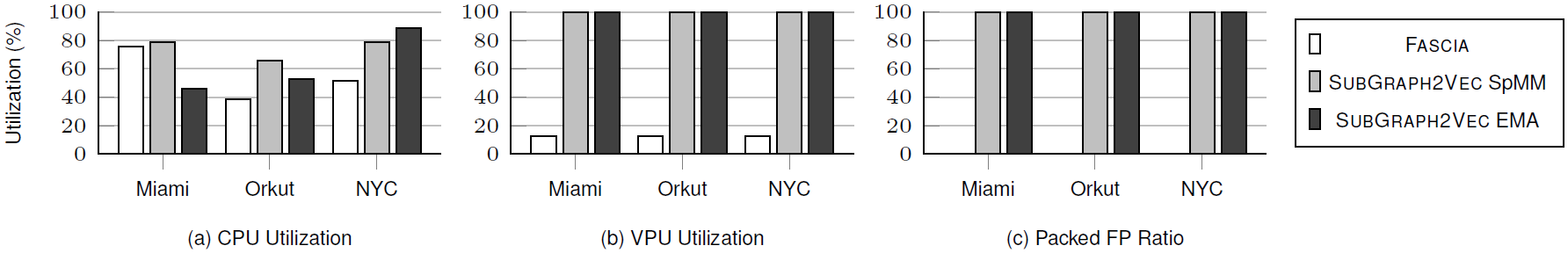}
    \caption{The hardware utilization on one Skylake node for template u12.}
    \label{fig:cpu-vpu-perf}
\end{figure*}
In the experiments, we use: 1) a single node of a dual-socket Intel(R) Xeon(R) CPU E5-2670 v3 
(architecture Haswell), 2) a single node of a dual-socket Intel(R) Xeon(R) Platinum 8160 CPU 
(architecture Skylake-SP) processors, and 3) a single node of Tesla V100 SXM2 paired with an 
Intel(R) Xeon(R) CPU E5-2630 v4.   
More details of the testbed hardware as well as the computation environment are released in the 
Artifact Description file.\par 
\begin{table*}[ht]
\renewcommand\arraystretch{0.80}
\centering
\caption{Datasets used in the experiments (K=$10^3$, M=$10^6$)}
\label{tab:datasets}
    \begin{tabular}{lllllll}
        \toprule
        Data  &  Vertices &  Edges &   Avg Deg &   Max Deg & Abbreviation & Source\\ 
        \midrule
        Human STRING10 & 10,971&214,298 &19.53 & 2009 & Human & Biology~\cite{hodzic2019combinatorial}\\
        EcoliGO-BP & 1,474 & 6,896 & 9.36 & 72 & Ecoli &  Biology~\cite{EColiNet}\\
        WI-2004 & 1,239 & 1,736 & 2.8 & 74 & Worm1 & Biology~\cite{Yeast}\\
        WI-2007 & 1,498 & 1,817 & 2.43 & 86 & Worm2 & Biology~\cite{Yeast}\\
        Combined-APMS & 1,622 & 9,070 & 11.18 & 127 & Yeast1 & Biology~\cite{Yeast}\\
        LC-multiple & 1,536 & 2,925 & 3.81 & 40 & Yeast2 & Biology~\cite{Yeast}\\
         Graph500 Scale=20  & 600K & 31M  & 48 & 67K & GS20 & Graph500~\cite{graphchallengemit}\\ 
         Graph500 Scale=21  & 1M & 63M  & 51 & 107K & GS21 & Graph500~\cite{graphchallengemit}\\ 
         Graph500 Scale=22   & 2M & 128M  & 53 & 170K & GS22 & Graph500~\cite{graphchallengemit}\\
        Miami &  2.1M & 200M &  49 &  10K & MI &  Social network~\cite{barrett_generation_2009}\\ 
        Orkut &  3M & 230M  & 76 & 33K  & OR &  Social network~\cite{snapnets}\\ 
        NYC   & 18M & 960M  & 54 & 429 & NY &  Social  network~\cite{yang_defining_2012}\\ 
        RMAT-1M & 1M & 200M & 201 & 47K  & RT1M & Synthetic data~\cite{chakrabarti_r-mat:_2004}\\
        RMAT(K=3) & 4M & 200M & 52 & 26K & RTK3 & Synthetic data~\cite{chakrabarti_r-mat:_2004}\\
        RMAT(K=5) & 4M & 200M & 73 & 144K & RTK5 & Synthetic data~\cite{chakrabarti_r-mat:_2004}\\
        RMAT(K=8) & 4M & 200M & 127 & 252K & RTK8 & Synthetic data~\cite{chakrabarti_r-mat:_2004}\\
        \bottomrule
    \end{tabular}
\end{table*}

\begin{table}[ht]
\renewcommand\arraystretch{0.80}
\centering
\caption{Execution time (s) of \ourimpl{} (S) versus \fasciaimpl{} (F) with increasing template sizes from U12 to U17.
Tests run on a Skylake node.}
\label{tab:ExecutionTime}
\setlength{\tabcolsep}{0.8mm}{
    \begin{tabular}{lllllllll}
        \toprule
        Dataset & Impl				&  u12 &  u13 &   u14 &   u15-1 & u15-2 & u16	&u17\\ 
        \midrule
        Miami& F 	& 163	&400	&944	&2663	&2435	&		&\\ 
        \rowcolor{Gray}
		Miami&S  	& 18	&38	&55	&160	&150	&		&\\  
        Orkut &F 	& 642	&2006	&4347	&1.5e4	&1.2e4	&		&\\ 
        \rowcolor{Gray}
		Orkut&S  	& 30	&67	&80	&238	&230	&		&\\ 
		
        RMAT 1M&F	& 1535	&5378	&1.2e4	&3.4e4	&3.2e4	&1.1e5		&3.8e5\\ 
        \rowcolor{Gray}
		RMAT 1M &S	& 16	&32	&34	&97	&97	&224		&573\\
		
        Graph500 20&F	& 132	&452	&923	&3379	&2679	&		&\\
        \rowcolor{Gray}
		Graph500 20&S	& 7	&14	&21	&63	&56	&		&\\ 
		
        Graph500 21&F	& 289	&1044	&2036	&7535	&5914	&		&\\ 
        \rowcolor{Gray}
		Graph500 21&S	& 12	&26	&36	&105	&102	&		&\\ 
		
        Graph500 22&F   & 764	&2814	&5477	&1.9e4	&1.6e4	&		&\\ 
        \rowcolor{Gray}
		Graph500 22&S   & 26	&53	&74	&220	&194	&		&\\ 
		
        RMAT K=3&F			& 1191	&4890	&9711	&3.0e4	&3.2e4	&		&\\ 
        \rowcolor{Gray}
		RMAT K=3&S			& 39	&110	&170	&377	&262	&		&\\ 
		
        RMAT K=5&F			& 2860	&9906	&2.0e4	&9.0e4	&5.4e4	&		&\\ 
        \rowcolor{Gray}
		RMAT K=5&S			& 29	&60	&82	&233	&240	&		&\\ 
		
        RMAT K=8&F			& 5620	&2.0e4	&3.3e4	&9.4e4	&8.5e4	&		&\\ 
        \rowcolor{Gray}
		RMAT K=8&S			& 25	&51	&67	&217	&234	&		&\\ 

        \bottomrule
    \end{tabular}
}
\end{table}
We use the following implementations.  
\begin{itemize}
    \item \textbf{\fasciaimpl{}} implements the graph traversal model of color-coding algorithm with multi-threading on a single CPU ~\cite{SlotaFastApproximateSubgraph2013}, which serves as a performance baseline.
    \item \textbf{\ourimpl{}} implements \ourimpl{} on CPU by using our in-house CSC-Split format with a SpMM kernel and eMA kernel 
    (threaded by OpenMP). It is the default implementation of \ourimpl{} and supports distributed systems.
    \item \textbf{\ourimpl{}-MKL} implements \ourimpl{} on a single CPU by using CSR based SpMV kernel from Intel MKL and 
    eMA kernel (threaded by OpenMP). It also supports distributed systems.
    \item \textbf{\ourimpl{}-cuSPARSE} \ourimpl{} on GPU by using CSR based SpMV kernel from NVIDIA cuSPARSE and eMA kernel (threaded by CUDA). Supports distributed systems by using the CSR format API from distributed mode of \ourimpl{}-MKL. 
\end{itemize}
Binaries on CPU are compiled by the Intel(R) C++ compiler for Intel(R) 64 target platform from Intel(R) Parallel Studio XE 2019, with compilation flags of ``-O3``, ``-xCore-AVX2", ``-xCore-AVX512", and the Intel(R) OpenMP. 
Binaries on GPU are compiled by CUDA release 9.1 (V9.1.85). 
The distributed binaries are compiled by Intel MPI 2019. 
We use, by default, a thread number equal to the physical core number of CPU, i.e., 48 threads on a Skylake node and 24 threads on a Haswell node. 
The threads are bind to cores with a spread affinity. 
For GPU, we use a thread block with a size of 1024 for the eMA kernel. 
For kernel invoked by Intel MKL and NVIDIA cuSPARSE, we use the default setup. 
We mainly use the \ourimpl{} to evaluate our work except for Section~\ref{sec:port-to-plat}, where \ourimpl{} with public library kernels are evaluated against \fasciaimpl{}.

\subsection{Overall Performance Improvement}
\label{sub:perf-improve}
We first examine the performance improvement of \ourimpl{} over the state-of-the-art \fasciaimpl{} on a Skylake node. The best performance we can obtain is by using a customized matrix format and SpMM kernel. 
Note that we scale the template size up to the memory limitation on our Skylake testbed for each dataset in Table~\ref{tab:ExecutionTime}. The reduction of execution time is significant, particularly for template sizes larger than 14. 
For instance, \fasciaimpl{} spends four days to process a million-vertex dataset RMAT-1M with template u17 while \ourimpl{} only spends 9.5 minutes. 
For relatively smaller templates such as u12, \ourimpl{} still achieves 10x to 100x of improvement on datasets Miami, Orkut, and RMAT-1M. \par 

In Table~\ref{tab:ExecutionTime}, we observe that the improvement is approximately proportional to the average degree of datasets. 
For instance, \ourimpl{} achieves 10x and 20x improvements on datasets Miami (average degree of 49) and Orkut (average degree of 76), respectively.
It implies that our optimization works better on dense graph network when compared to \fasciaimpl{}.\par

The three Graph500 datasets in Table~\ref{tab:datasets} have comparable average degrees but growing vertex number and edge number. 
For the same template, \ourimpl{} obtains similar improvements over \fasciaimpl{} across the three datasets implying that \ourimpl{} has a scalable performance improvement with respect to the dataset size. \par  

Finally, we compare RMAT datasets with increasingly skewed degree distribution, which causes a thread-level workload imbalance.
The results show that \ourimpl{} has comparable execution time regardless of the degree distribution.
On the contrary, \fasciaimpl{} spends significantly (2x to 3x) more time on datasets with skewed degree distribution.

\subsection{Benefit of System Design}
\label{sub:perf-benefit-sys-design}


\begin{figure}[ht]
    \centering
    \subfloat[Non-Vectorized versus Vectorized]{\includegraphics[width=0.90\linewidth]{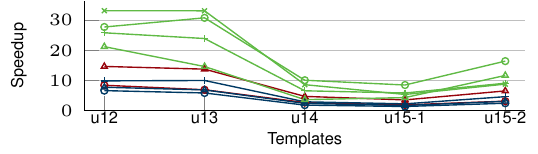}}
    \vfill
    \subfloat[Overall Performance Improvement ]{\includegraphics[width=0.90\linewidth]{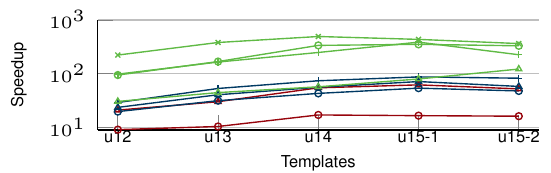}}
    \vfill
    \includegraphics[width=0.70\linewidth]{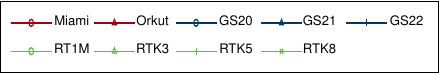}
    \caption{The performance improvement brought by vectorization and the overall performance improvement. The Scaling tests run on a Skylake node.}
    \vspace{-3ex}
    \label{fig:vecspeedup}
\end{figure}

To evaluate the benefit of our system design, we remove the optimization in Section~\ref{sec:sys-design} while only applying multi-threading in both stages of Algorithm~\ref{alg:prune-redundancy} as a baseline in Figure~\ref{fig:vecspeedup}. 
We observe that on Miami dataset, \ourimpl{} obtains 8x speedup by average and up to 30x for some templates. 
The reason behind that is a significant improvement in hardware utilization, which is evaluated from: 1) The CPU and VPU utilization, 2) An overall efficiency by using the roofline model.

\subsubsection{\textbf{CPU and VPU Utilization}}
Figure~\ref{fig:cpu-vpu-perf}(a) first compares the CPU utilization defined as the average number of concurrently running physical cores. 
For Miami, \fasciaimpl{} achieves 60\% of CPU utilization. However, the CPU utilization drops down below 50\% on Orkut and NYC. Conversely, SpMM kernel keeps a high CPU utilization from 65\% to 78\% for all datasets. The eMA kernel has a growing CPU utilization from Miami (46\%) to NYC (88\%). 
We have two explanations: 
1) the SpMM kernel splits and regroups the nonzero entries by their row IDs, which mitigates the imbalance of nonzero entries among rows;
2) the eMA kernel has its computation workload for each column of $\mathbf{M}_{s,a}, \mathbf{M}_{s,p}$ evenly dispatched among threads. \par
Secondly, we examine the code vectorization in Figure~\ref{fig:cpu-vpu-perf}. 
VPU in a Skylake node is a group of 512-bit registers. 
The scalar instruction also utilizes the VPU, but it cannot fully exploit its 512-bit length.
Figure~\ref{fig:cpu-vpu-perf} refers to the portion of instructions vectorized with a full vector capacity. 
For all of the three datasets, \fasciaimpl{} only has 6.7\% to 12.5\% VPU utilization, implying that the codes are not vectorized.
While for SpMM and eMA kernels of \ourimpl{}, the VPU utilization is 100\%. 
A further metric of packed float point instruction ratio (Packed FP) justifies the implication that \fasciaimpl{} has zero vectorized instructions, but \ourimpl{} has all of its float point operations vectorized. 

\subsubsection{\textbf{Roofline Model}}
\label{sub:roofline}
\begin{figure}[ht]
 \centering
 \includegraphics[width=0.85\linewidth]{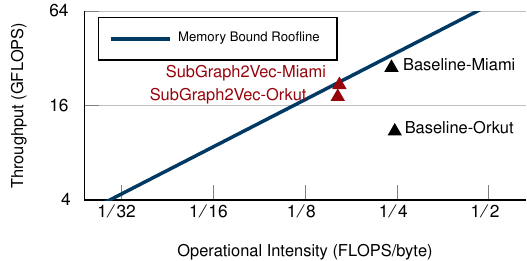}
 \caption{Apply roofline model to \fasciaimpl{} and \ourimpl{}. Dataset Miami, Orkut for template u15-1. Tests are done on a Skylake node.
}
\label{fig:roofline}
\end{figure}

The roofline model in Figure~\ref{fig:roofline} reflects hardware efficiency. The horizontal axis is the operational intensity (FLOP/byte), and the vertical axis refers to the measured throughput performance (FLOP/second). 
The solid roofline is the maximal performance the hardware can deliver under a certain operational intensity.
Because of the low operational intensity, the performance of \fasciaimpl{} and \ourimpl{} are bounded by the memory bandwidth, and we consider it as a memory-bound roofline. 
For a relatively small dataset like Miami, both of \fasciaimpl{} and \ourimpl{} are close to the memory-bound roofline because the data can be fit into the 33 MB L3 cache. 
For dataset Orkut, whose data size is beyond the capacity of L3 cache, \ourimpl{} is much closer to the memory-bound roofline than that of \fasciaimpl{} because of its regular and vectorized memory access pattern.
 \subsubsection{\textbf{Memory Bandwidth and Cache Usage}}
 \label{sub:bd-cache}
 \begin{table}[ht]
 \renewcommand\arraystretch{0.80}
     \centering
     \caption{Memory and Cache Usage of \fasciaimpl{}, SpMM, and eMA of 
     \ourimpl{} on a Skylake Node}
     \setlength{\tabcolsep}{0.5mm}{
     \begin{tabular}{lllll}
     \toprule
     Orkut & Bandwidth & L1 Miss Rate & L2 Miss Rate & L3 Miss Rate  \\
     \midrule
     \fasciaimpl{} & \textbf{8 GB/s} & 9.6\% & 5.3\% & \textbf{46\%} \\ 
     SpMM & \textbf{59.5 GB/s} & 6.7\% & 42.8\% & \textbf{45\%} \\ 
     eMA & \textbf{116 GB/s} & 0.32\% & 22.2\% & \textbf{9.0\%} \\ 
     \toprule
     NYC & Bandwidth & L1 Miss Rate & L2 Miss Rate & L3 Miss Rate  \\
     \midrule
     \fasciaimpl{} &  \textbf{7 GB/s} & 2.4\% & 8.1\% & \textbf{87\%} \\ 
     SpMM & \textbf{96 GB/s} & 7.7\% & 76\% & \textbf{74\%} \\ 
     eMA & \textbf{122 GB/s} & 0.1\% & 99\% & \textbf{14.8\%} \\ 
     \bottomrule
     \end{tabular}
     }
     \label{tab:exp:mem-cache}
 \end{table}
In Table~\ref{tab:exp:mem-cache}, we compare SpMM and eMA of \ourimpl{} to \fasciaimpl{}. 
It shows that the eMA kernel has the highest bandwidth value around 110 GB/s for the three datasets, which is due to the highly vectorized codes and regular memory access patterns. 
The data is prefetched into cache lines, which mitigates the cache miss rate as low as 0.1\%. 
The SpMM kernel also enjoys a decent bandwidth usage of around 70 to 80 GB/s by average when compared to \fasciaimpl{}. 

\subsection{Parallelization on a Single Node}
\begin{figure}[ht]
 \centering
 \includegraphics[width=0.85\linewidth]{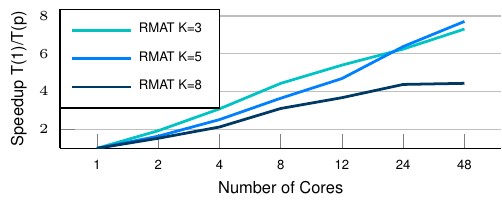}
 \vspace{-1ex}
 \caption{Strong scaling test of RMAT datasets with increasing cores on a Skylake node.}
\label{fig:thdscale}
\end{figure}

We perform a strong scaling test using up to 48 threads on Skylake node in Figure~\ref{fig:thdscale}. We choose RMAT generated datasets with increasing skewness parameters of $K=3, 5, 8$. 

As the performance is bounded by memory, which has 6 memory channels per socket, we have a total of 12 memory channels on a Skylake node that bounds the thread scaling. 
Eventually, \ourimpl{}  obtain a 7.5x speedup at 48 threads when $K=3$. 
When increasing the skewness of datasets to $K=5,8$, the thread scalability of  \ourimpl{}  drop down because the skewed data distribution brings workload imbalance. 
\begin{figure*}[ht]
    \centering
    \includegraphics[width=0.85\textwidth]{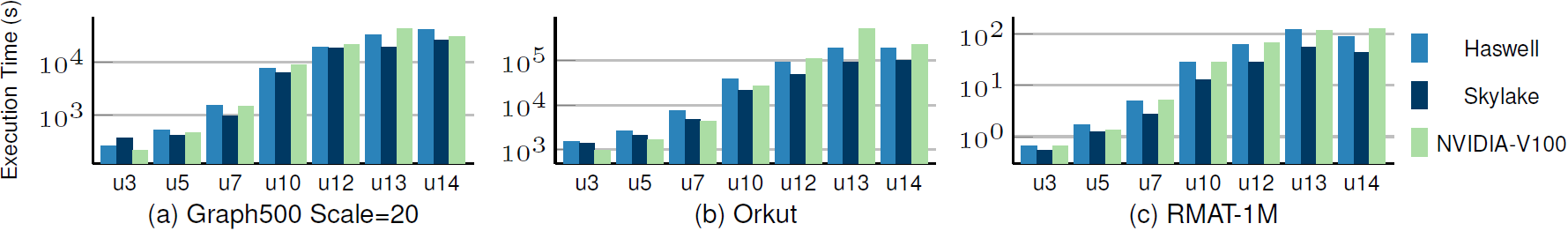}
     \caption{Execution Time of \ourimpl{} on three platforms. On Haswell and Skylake nodes, 
     we use CSR based SpMV kernel from Intel MKL; On Volta GPU V100, 
     we use CSR based SpMV kernel from NVIDIA cuSPARSE}
    \label{fig:cross-plat}
\end{figure*}
 
\subsection{From Single Node to Distributed System}
\label{sub:single-to-distri}

\begin{figure}[ht]
    \centering
    \includegraphics[width=0.85\linewidth]{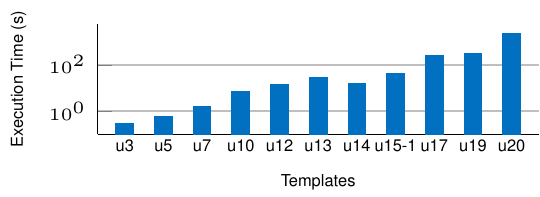}
     \vspace{-1ex}
     \caption{Scaling up the templates up to u20 by distributed \ourimpl{} on 16 Haswell nodes.}
    \label{fig:scalermat-tp}
    \centering
    \vspace{1ex}
    \includegraphics[width=0.85\linewidth]{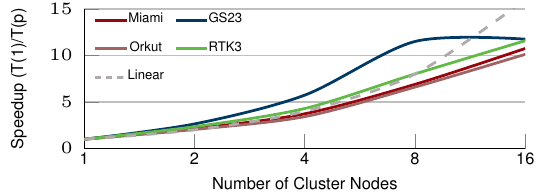}
     \vspace{-1ex}
     \caption{Strong scaling test on distributed \ourimpl{}. Four datasets with template u14 running on 16 Haswell nodes.}
    \label{fig:distri_sub2vec_sscaling}
\end{figure}

A distributed \ourimpl{} extends the memory capacity of a single node that enables to run larger templates.
As an example, dataset RT1M in Table~\ref{tab:ExecutionTime} can only run templates with size up to u17 on a single node.
However, we can scale up the template size to u20 with a cluster of 16 nodes as shown in Figure~\ref{fig:scalermat-tp}.\par
Our distributed \ourimpl{} has a good strong scalability, which even achieves super linear speedup in Figure \ref{fig:distri_sub2vec_sscaling} from 1 node to 8 nodes. 
According to the analysis in Section~\ref{sub:roofline}, \ourimpl{} is memory bounded, and increasing the number of nodes scales out not only computation resources but also memory bandwidth and cache resources.
Having less data on each node can increase the percentage of data held by the last level of the CPU cache. 
\subsection{Portability to Other Platforms}
\label{sec:port-to-plat}

Hardware platforms such as NVIDIA GPU already have highly-optimized public libraries of linear algebra kernels.
For SpMV operation, we have the \emph{mkl\_sparse\_s\_mv} kernel from Intel MKL library on CPU and the \emph{cusparseScsrmv} kernel from NVIDIA cuSPARSE library on GPU. 
For eMA kernel, we can use a combination of \emph{vsMul} and \emph{vsAdd} kernels from Intel MKL or hand-implement such kernels whenever the kernel is absent because of its simplicity. 
Hence, we have ported \ourimpl{} to GPU by keeping the CPU codes other than SpMV and eMA on the host side while invoking cuSPARSE and CUDA kernels for the two linear algebra operations.

In Figure~\ref{fig:cross-plat}, we port the performance of \ourimpl{} to three platforms by using CSR-SpMV libraries kernels. When the template size is small, \ourimpl{}-cuSPARSE has comparable or even better performance than \ourimpl{}-MKL. However, the performance of \ourimpl{}-cuSPARSE drops down when the template size grows up. 
As NVIDIA-V100 only has 16GB of device memory, it is probable that the large memory footprint of $M_s$ brought by large template size cannot fit into the device memory, and the bi-directional data transfer between the host and device memory compromises the performance of \ourimpl{}-cuSPARSE.
Nevertheless, both of Intel MKL and NVIDIA cuSPARSE are not open-sourced, and we cannot conclude on their performance gap. 
Also, the three hardware platforms have different theoretical peak performances and memory bandwidths. This result is only meant to demonstrate the portability of our \ourimpl{} across hardware platforms.

\subsection{Error Discussion}
\label{sub:error-discussion}
We implement the standard color coding algorithm that Alon et al. \cite{AlonBiomolecularNetworkMotif2008} prove to run at most $N$ iterations to control approximation quality as in Algorithm ~\ref{alg:sequential}.
In practice, the subgraph counting with color-coding requires only 100 iterations for a 7 node template on H.pylori with an error of less than 1\% in FASCIA \cite{slota_parallel_2015}. 
\ourimpl{} with its pruning and vectorization optimization only differs from the \fasciaimpl{} due to the restructuring of the computation from Algorithm~\ref{alg:seq-dp} to Algorithm~\ref{alg:prune-redundancy}. 
It should give identical results with exact arithmetic in Equation~\ref{eq:transform_traversal}. 

\begin{figure}[ht]
    \centering
    \includegraphics[width=0.95\linewidth]{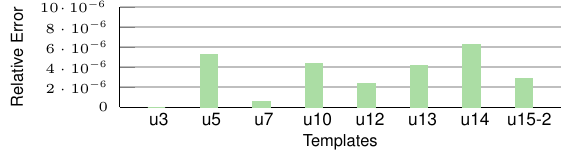}
    \vspace{-1ex}
     \caption{Relative error on dataset Graph500 Scale=20.}
    \label{fig:error}
\end{figure}
However, when dealing with large graphs, the counted value will exceed the range of integer variables. 
As a consequence, both \fasciaimpl{} and our \ourimpl{} use 32-bit floating-point numbers to avoid overflow. 
Hence, slightly different results are observed between \fasciaimpl{} and \ourimpl{} due to the rounding error consequent from floating-point arithmetic operations. 
Figure~\ref{fig:error} reports such relative errors between \ourimpl{} and \fasciaimpl{} in the range of $10^{-6}$ across all the tests on a Graph500 GS20 dataset with increasing template sizes, which is negligible.

\section{Conclusion}
\label{sec:conclusion}
 
In this paper, we fully vectorize a sophisticated algorithm of subgraph analysis, and the novelty is a co-design approach with pattern identification of linear algebra kernels that leverage hardware vectorization of Intel CPU and NVIDIA GPU architectures.
The overall performance achieves significant improvements over the state-of-the-art work by orders of magnitude by average and up to 660x (RMAT1M with u17) within a shared-memory multi-threaded system.
The distributed memory system runs large tree subgraphs with sizes up to 20. 

This work demonstrates that fundamental algorithms, such as subgraph mining, could be efficiently explored.
More significantly, by implementing it with hardware vectorization, we point a direction where scaling performance of complex graph applications with random access to the vast memory region and dynamic programming workflow is possible and can be done with sparse linear algebra kernels in contrast to conventional graph traversal.
An interesting future work will be exploring machine learning as the irregularity of memory access remains a roadblock to improve hardware utilization.
It will lead to new research and improvements on portability to other emerging hardware accelerators and heterogeneous architectures.  

\section{Acknowledgement}
\label{sec:acknowledgement}
We gratefully acknowledge the support from NSF CIF21 DIBBS 1443054: Middleware and High Performance Analytics Libraries for Scalable Data Science, Science and NSF EEC 1720625: Network for Computational Nanotechnology (NCN) – Engineered nanoBio node, NSF OAC 1835631 CINES: A Scalable Cyberinfrastructure for Sustained Innovation in Network Engineering and Science, and Intel Parallel Computing Center (IPCC) grants. We would like to express our special appreciation to the FutureSystems team.
\bibliographystyle{ieeetr}
\bibliography{sample-base}
\end{document}